\documentclass[epjST]{svjour}

\usepackage{mathptmx}
\usepackage{amsmath}
\usepackage{graphicx}
\usepackage{citesort}
\usepackage{color}
\usepackage{float}
\usepackage{placeins}
\usepackage{amsmath, amssymb}

\begin{document}
\title{Initial correlated states for the Generalized Kadanoff--Baym Ansatz without adiabatic switching-on of interactions in closed systems }
\author{Miroslav Hopjan\inst{1,2}\fnmsep\thanks{\email{miroslav.hopjan@teorfys.lu.se}} \and Claudio Verdozzi\inst{1,2}\fnmsep\thanks{\email{claudio.verdozzi@teorfys.lu.se}}}
\institute{Department of Physics, Lund University, PO Box 118, 221 00 Lund, Sweden  \and European Theoretical Spectroscopy Facility (ETSF)}
\abstract{
We reconsider the Generalized Kadanoff--Baym Ansatz
(GKBA) approximation for non-equilibrium Green's functions 
and extend it to self-consistently define an equilibrium
correlated (within GKBA) state in closed systems. The advantage of
the proposed prescription is to avoid the preparation of the initial equilibrium correlated 
state via adiabatic switching-on of the correlations. A simple model system, 
namely a Hubbard-dimer, is used to illustrate aspects of the computational
implementation and performance of the new scheme.
} 
\maketitle
\section{Introduction}
\label{intro}
The description of time-dependent processes in non-equilibrium quantum systems can be 
formulated within the Nonequilibrium Green's function (NEGF) formalism, i.e. in terms of contour 
Green's function $G(z,z')$, where $z$ and $z'$ are generalized times running over 
Martin--Schwinger--Keldysh contour $\gamma$~\cite{KBE,Keldysh,BalzBon,StefLeeu,Hopjan14,commentKBE}.  
The Green's function can be propagated from its initial equilibrium value (\textit{the initial equilibrium
state}) according to the Kadanoff--Baym equations (KBE) \cite{KBE}. In the last two decades,
there have been several works devoted to the numerical implementation of the KBE; among the applications, we mention simple models of 
atoms or molecules \cite{Dahlen2007,Balzer2008}, quantum dots \cite{Myohanen2008} and lattice systems
\cite{Friesen2009,Friesen2010,Hopjan16,Schlunzen2016,Schlunzen2016a,Schlunzen2017}. The implementation
of the KBE requires to propagate two-time integro-differential equations. Due to memory costs, only limited 
systems sizes (or propagation times) can be reached \cite{Schlunzen2016a}.
In the KBE method there are 2 ways to prepare the equilibrium correlated state. The first way
is based on the Martin--Schwinger--Keldysh contour with the imaginary time appendix, representing the
equilibrium state \cite{StefLeeu,Friesen2009,Friesen2010,Danielewicz,Pavlyukh2017,Pavlyukh2018}. The second way is via the so-called adiabatic switching procedure
\cite{Keldysh}, where one starts the evolution from an uncorrelated (noninteracting or Hartree--Fock) state 
and adiabatically switches on the correlation during the evolution, see e.g. Ref. \cite{Schlunzen2016a}
for numerical details. Both methods gives the same results, as numerical comparisons show \cite{commentKBE,Schlunzen2016a},
but the latter has larger memory costs \cite{Schlunzen2016a}.\\

To overcome the size and the propagation-time limitations of the KBE, we can propagate directly a 
transport equation for the density matrix $\rho(t)$.  The density matrix is the equal-time lesser Green's
function $\rho_{}(t)= -i G^{<}_{}(t,t)$, where the function $G^{<}(t,t')$ follows from the contour Green's
function $G(z,z')$ by setting $z=t$ on the forward branch and $z'=t'$ on the backward branch of the 
Martin--Schwinger--Keldysh contour \cite{StefLeeu} (in this way we lose access to the spectral features).
The transport equation for $\rho$ can be formally derived from the KBE, and with help of so-called Generalized 
Kadanoff--Baym Ansatz (GKBA) \cite{Lipavsky1986,Bonitz1988,Velicky2008} we can close the transport equations. 
In recent years there has been an increasing number of implementation of the GKBA transport equations in 
inhomogeneous systems \cite{Schlunzen2016a,Schlunzen2017,Hermanns2012,Hermanns2014,Latini2014,BarLev16,Spicka2017,Hopjan2017,Perfetto2018,VinasBostrom2018,Spicka2018}. 
In these studies, the initial equilibrium correlated density matrix (\textit{the initial equilibrium state}) was 
typically prepared via adiabatic switching of the electronic correlations. So far an alternative way to access
the initial equilibrium correlated state is lacking. The problems is addressed here, where we propose a method,
to obtain the initial equilibrium density matrix directly, i.e. by avoiding the adiabatic switching.

\section{Generalized Kadanoff Baym Ansatz - overview}
The exact transport equation for the density matrix $\rho$, which can be derived from the Kadanoff--Baym equations (KBE),
reads in matrix form 
\begin{align}
\partial_{t}{\rho}_{}(t)+i[h_{HF}[{\rho}_{}(t)],{\rho}_{}(t)]=-(I^{<}_{}(t,t)+{h.c.}),
\end{align}
where $h_{HF}$ is the Hartree-Fock term and the collision term $I^{<}_{}$ reads 
\begin{align}
I^{<}_{}(t,t)=\int_{-\infty}^{t}d\bar{t} (\Sigma^{<}_{}[G](t,\bar{t})G^{A}_{}(\bar{t},t)+\Sigma^{R}_{}[G](t,\bar{t})G^{<}_{}(\bar{t},t)).
\end{align}
Here $<$ and $R$ respectively label the lesser and retarded components of the Green's function and self-energy \cite{Langreth}. 
The self-energy can be composed from the correlation part and from the embedding part $\Sigma^{}_{}[G]=\Sigma^{}_{\rm corr.}[G]+\Sigma^{}_{\rm emb.}$. The correlation self-energy is a 
functional of the Green's function, while for closed systems, as considered here, $\Sigma^{}_{\rm emb.}=0$. The exact transport equation is not a closed equation for $\rho$. To close 
the equation we can use so-called reconstruction equations \cite{Lipavsky1986}:
\begin{eqnarray}\label{construction}
&{G}^{<}_{}(t,t')=-{G}^{R}_{}(t,t')\rho_{}(t')+\rho_{}(t){G}^{A}_{}(t,t')
\dots,\\\label{construction2}
&{G}^{>}_{}(t,t')={G}^{R}_{}(t,t')(1-\rho_{}(t'))-(1-\rho_{}(t)){G}^{A}_{}(t,t')+\dots.
\end{eqnarray}
where the dots stands for terms of higher order in $\rho$ and ${G}^{R/A}_{}$. The sum of all terms of such expansion gives back the 
solution of the full KBE \cite{Lipavsky1986}. In such case the Green's function components satisfy the relation 
\begin{eqnarray}\label{relation}
{G}^{>}(t,t')-{G}^{<}(t,t')={G}^{R}(t,t')-{G}^{A}(t,t').
\end{eqnarray}
and the $\rho$ from the transport equation connects to the Green's function from the KBE  via $\rho(t)=-iG^{<}(t,t)$.\\

In the Generalized Kadanoff--Baym Ansatz (GKBA) the reconstruction-equation expansion is truncated after the first order, i.e.
\begin{eqnarray}\label{construction}
&\tilde{G}^{<}_{}(t,t')=-\tilde{G}^{R}_{}(t,t')\rho_{}(t')+\rho_{}(t)\tilde{G}^{A}_{}(t,t'),\\\label{construction2}
&\tilde{G}^{>}_{}(t,t')=\tilde{G}^{R}_{}(t,t')(1-\rho_{}(t'))-(1-\rho_{}(t))\tilde{G}^{A}_{}(t,t'),
\end{eqnarray}
where we denote the approximate GKBA Green's function by $_\text{\LARGE \textasciitilde}$. The components of the GKBA 
Green's function, according to their definition, automatically satisfy an analog of Eq. \eqref{relation}:
\begin{eqnarray}\label{relation2}
\tilde{G}^{>}_{}(t,t')-\tilde{G}^{<}_{}(t,t')=\tilde{G}^{R}_{}(t,t')-\tilde{G}^{A}_{}(t,t').
\end{eqnarray}
By construction, the relation  $\rho=-i\tilde{G}^{<}$ is also identically satisfied. Another important aspect to be noted is that the lesser and greater GKBA Green's functions are fully characterized only when
the retarded and advanced components in Eq. \eqref{construction} are specified. The equation of motion for 
such an auxiliary Green's function reads
\begin{equation}
(i\partial_{t}-h^{}_{HF}[\rho](t))\tilde{G}^{R/A}_{}(t,t')=\delta(t-t')+\int_{t'}^{t}d\bar{t} \tilde{\Sigma}^{R/A}_{}(t,\bar{t})\tilde{G}^{R/A}_{}(\bar{t},t').
\end{equation}
Here the level of approximation of $\tilde{G}$ is given by the auxiliary self-energy $\tilde{\Sigma}$. Therefore there is a freedom in its choice \cite{Lipavsky1986,Hermanns2012} and in most of the practical implementations the auxiliary self-energy is different from self-energy in the transport equation $\Sigma\neq\tilde{\Sigma}$. Typically we 
choose $\tilde{\Sigma}^{R}_{}(t,\bar{t})=\tilde{\Sigma}^{R}_{}(t)\delta(t-\bar{t})$ local in time, in order 
to minimize the computational costs. In closed systems, auxiliary retarded and advanced 
Green's functions can be constructed from the density matrix $\rho$ at the Hartree-Fock (HF) level \cite{Hermanns2012}, i.e. $\tilde{\Sigma}^{R}_{}(t,\bar{t})=0$.\\

The GKBA auxiliary Green's functions are then used to close the transport equation, since they
approximate the scattering integral $I[G]\rightarrow I[\tilde{G}]$ explicitly as
\begin{eqnarray}
I^{<}_{}(t,t)\approx \int_{-\infty}^{t}d\bar{t} (\Sigma^{<}_{}[\tilde{G}^{}](t,\bar{t})\tilde{G}^{A}_{}(\bar{t},t)+\Sigma^{R}_{}[\tilde{G}^{}](t,\bar{t})\tilde{G}^{<}_{}(\bar{t},t)),
\end{eqnarray}
where the auxiliary GKBA functions are also used in the construction of correlations selfenergies, see Fig.\ref{chart} i). The interesting 
property of the GKBA is the fulfilment of the relation for the selfenergies
\begin{eqnarray}\label{relation3}
\Sigma^{>}_{}[\tilde{G}^{}](t,t')-\Sigma^{<}_{}[\tilde{G}^{}](t,t')=\Sigma^{R}_{}[\tilde{G}^{}](t,t')-\Sigma^{A}_{}[\tilde{G}^{}](t,t'),
\end{eqnarray}
even if the auxiliary GKBA Green's functions $\tilde{G}$ are used for the construction. The formal proof of this 
relation is similar to the proof of relation for the full Green's function dependence. The relation 
$\Sigma^{>}-\Sigma^{<}=\Sigma^{R}-\Sigma^{A}$ holds since each component of the self-energy is constructed according
to the Langreth--Wilkins rules \cite{Langreth}. To derive the latter, Eq. \eqref{relation} is 
usually used. Here the GKBA analog Eq. \eqref{relation2} can be used instead to derive similar rules. For the model system 
used below to test our approach, we have verified Eq. \eqref{relation3} analytically and numerically (not shown here).

\section{Generalized Kadanoff Baym Ansatz - equilibrium correlated state}

In the GKBA approximation the commonly used strategy is the adiabatic switching, where an 
uncorrelated Hartree--Fock density matrix is constructed and then propagated according to the adiabatic 
switching protocol \cite{Schlunzen2016a,Hermanns2012,Latini2014} from $t=-\infty$ to $t=0$.
After such preparation, one obtains a correlated density matrix which fulfils a 
steady state equilibrium version of the transport equation
\begin{align}\label{ssgkba}
i[h_{HF}^{}[\rho_{\rm eq.}],\rho_{\rm eq.}]=-(I^{<}(0,0)+h.c.)
\end{align}
where the scattering integral reads
 \begin{align}
 I^{<}_{}(0,0)={\int_{-\infty}^{0}}d\bar{t} (\Sigma^{<}_{}[\tilde{G}^{}](0,\bar{t})\tilde{G}^{A}_{}(\bar{t},0)+\Sigma^{R}_{}[\tilde{G}^{}](0,\bar{t})\tilde{G}^{<}_{}(\bar{t},0)).
 \end{align}
One can wonder if this equation can be used to directly generate the equilibrium correlated density matrix.
The answer is negative; some elements of such a matrix equation give the trivial relation $0=0$. 
This happens in closed systems, where the collision term contains only the correlation self-energy.

Thus, in closed systems, such set of equations Eq. \eqref{ssgkba} is under-determined  and we have 
to search for another way to determine the equilibrium density ${\rho}_{\rm eq.}$. 
The prescription proposed in this paper is to instead search for a equilibrium state Green's function 
$G^{<}_{\rm eq.}(\omega)$ which gives the density matrix via 
\begin{align}
{\rho}_{\rm eq.}=-i\int \frac{d\omega}{2\pi} G^{<}_{\rm eq.}(\omega)=-i\int \frac{d\omega}{2\pi} (-2i){\rm Im} G^{R}_{\rm eq.}(\omega)f(\omega). 
\label{uno}
\end{align}
where $f$ is the Fermi-Dirac function. We stress that the equilibrium state Green's function \textit{is not} the GKBA function. In the case
of the GKBA Green's function $\tilde{G}^{<}$ the relation ${\rho}_{\rm eq.}=-i\int \frac{d\omega}{2\pi} \tilde{G}^{<}_{\rm eq.}(\omega)$
is only a \textit{tautology}. In order to find the desired equilibrium function we have to extend the GKBA
approximation from the transport equation for $\rho$ to the full KBE, as discussed in the following section.\\

Before moving to the next section, it is important to stress that with our procedure as presented in Eq.\eqref{uno} and
detailed below, one is solving for the equilibrium density matrix $\rho_{\rm eq.}$
but not for $I^{<}(0,0)$. However, as recently shown in Ref. \cite{Karlsson2018}, the scattering integral $I^{<}(0,0)$ will
be needed if one wants to start the propagation at time $t=0$ using $\rho_{\rm eq.}$ as the starting point. This 
augmented term persists at subsequent times, in the form of 
\begin{align}
 I^{<}_{}(t,t)={\int_{-\infty}^{0}}d\bar{t} (\Sigma^{<}_{}[\tilde{G}^{}](t,\bar{t})\tilde{G}^{A}_{}(\bar{t},t)+\Sigma^{R}_{}[\tilde{G}^{}](t,\bar{t})\tilde{G}^{<}_{}(\bar{t},t)),
 \end{align}
and prevents spurious oscillations of the density \cite{Karlsson2018}.

\section{Extended Generalized Kadanoff Baym Ansatz}
\begin{figure}[t!]
\centering
\includegraphics[width=0.9\textwidth]{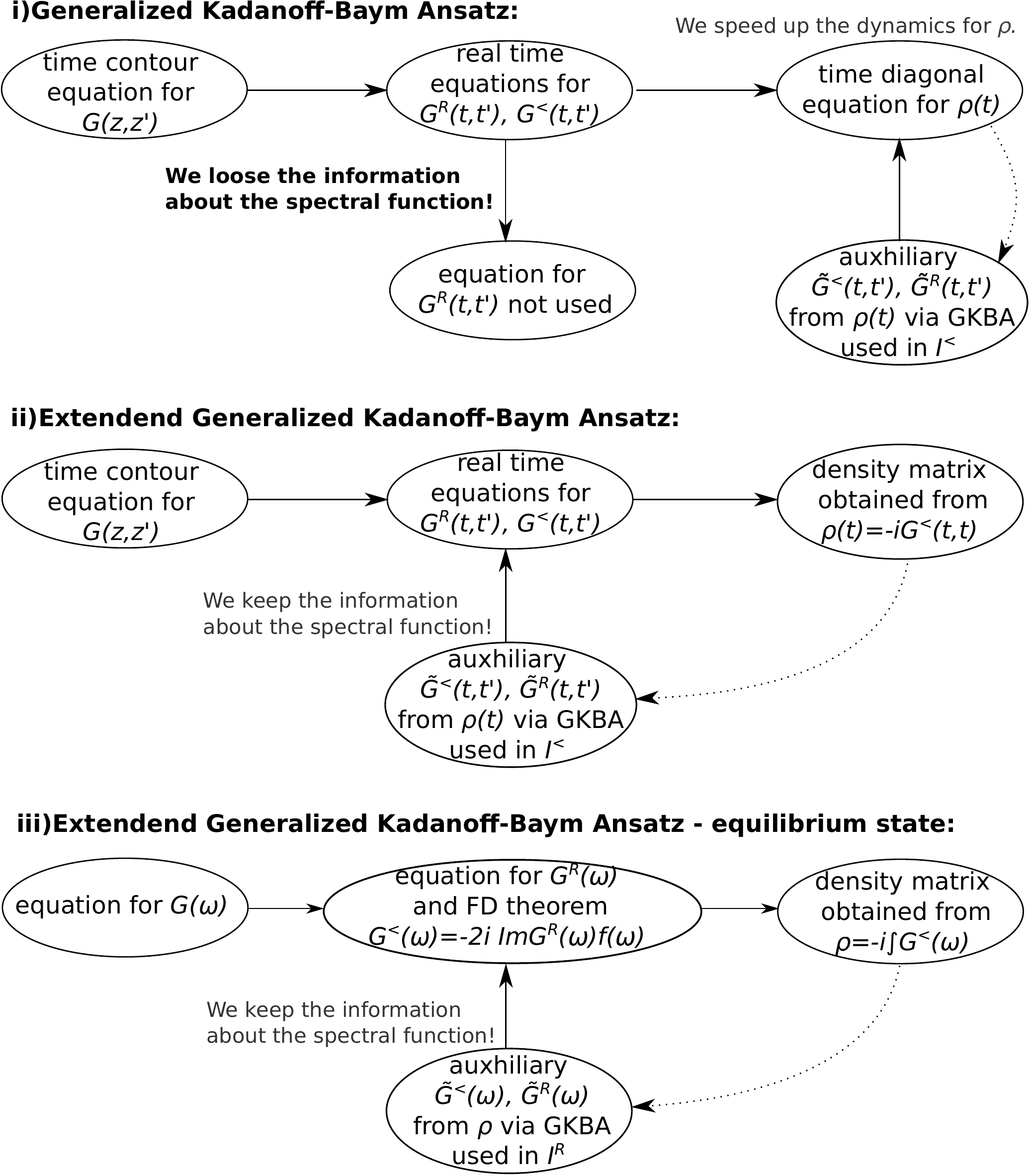}
\caption{A flowchart showing the loss of the spectral information within the Generalized 
Kadanoff--Baym Ansatz (GKBA) applied in the collision integral $I^{<}$ of the time diagonal transport
equation, panel i). The information will be restored if the Ansatz is applied in the real time 
equations for  $G^{R}$ and  $G^{<}$, panel ii). Such extended Generalized Kadanoff--Baym Ansatz (eGKBA)
is used in equilibrium calculations, where the equation for $G^{R}$ is to be solved with the help of
the fluctuation-dissipation (FD) theorem, panel iii). The dotted arrows indicate the closure of the equations.}
\label{chart}
\end{figure}

The GKBA is formulated for a time-diagonal transport equation and it aims directly at the time 
evolution of the density matrix. In this way we loose the part of information contained in the 
Green's function, notably the spectral function, which may be used for the construction of the 
equilibrium state. Hence we wish now to investigate what will happen if we extend the GKBA approximation 
to the off-diagonal equation. This corresponds to use $I[G]\rightarrow I[\tilde{G}]$ in the
collision part of the KBE equations for $G^{\lessgtr}$, see Fig.\ref{chart} ii). \\

In this extended Generalized Kadanoff--Baym Ansatz (eGKBA for simplicity), we approximate the collision part of the KBE (note the off-diagonal structure $t,t'$)
\begin{eqnarray}
(i\partial_{t}-h^{}_{HF}[\rho](t))G^{\lessgtr}_{}(t,t')=\int_{-\infty}^{t}d\bar{t} (\Sigma^{\lessgtr}_{}[\tilde{G}^{}](t,\bar{t})\tilde{G}^{A}_{}(\bar{t},t')+\Sigma^{R}_{}[\tilde{G}^{}](t,\bar{t})\tilde{G}^{\lessgtr}_{}(\bar{t},t')).
\end{eqnarray}
This differential equations can be formally solved by integration starting from the initial uncorrelated Hartree--Fock equilibrium state \cite{Keldysh}, schematically written
\begin{equation}\label{gkbaglg}
G^{\lessgtr}=G^{\lessgtr}_{HF} + G^{\lessgtr}_{HF} \Sigma^{A}[\tilde{G}] \tilde{G}^{A} + G^{R}_{HF} \Sigma^{\lessgtr} [\tilde{G}] \tilde{G}^{A} + G^{R}_{HF} \Sigma^{R}[\tilde{G}] \tilde{G}^{\lessgtr}.
\end{equation}
This solution is our desired Green's function, and its components have to obey equation 
\eqref{relation}. Below we show that the corresponding evolution of the retarded (advanced) Green's function 
must then obey to
\begin{eqnarray}
(i\partial_{t}-h^{}_{HF}[\rho](t))G^{R/A}_{}(t,t')=\delta(t,t')+\int_{t'}^{t}d\bar{t} \Sigma^{R/A}_{}[\tilde{G}^{}](t,\bar{t})\tilde{G}^{R/A}_{}(\bar{t},t').
\end{eqnarray}
This is formally solved by integration, starting from the initial uncorrelated Hartree--Fock equilibrium state, and schematically written as
\begin{equation}\label{gkbagra}
G^{R/A}=G^{R/A}_{HF} + G^{R/A}_{HF} \Sigma^{R/A}[\tilde{G}] \tilde{G}^{R/A}.
\end{equation}
We note that this equation is not a standard Dyson equation, as $G^{R/A}$ does not appear on the right hand site of the equation.\\

{\it Consistency check.--} To check that Eq. \eqref{relation} is satisfied by the evolution integral equations Eq. \eqref{gkbaglg} and Eq. \eqref{gkbagra}, we start by the 
difference of the left hand sides of Eq. \eqref{gkbaglg}, so we have
\begin{eqnarray} \nonumber
G^{>}-G^{<}=(G^{>}_{HF}-G^{<}_{HF}) + (G^{>}_{HF}-G^{<}_{HF}) \Sigma^{A} \tilde{G}^{A} \\
+ G^{R}_{HF} (\Sigma^{>}-\Sigma^{<}) \tilde{G}^{A} + G^{R}_{HF} \Sigma^{R} (\tilde{G}^{>}-\tilde{G}^{<}).
\end{eqnarray}
Then, using that ${G}^{>}_{HF}-{G}^{<}_{HF}={G}^{R}_{HF}-{G}^{A}_{HF}$, and also the relations Eq. \eqref{relation2} and Eq. \eqref{relation3} discussed above we get
\begin{eqnarray} \nonumber
G^{>}-G^{<}=(G^{R}_{HF}-G^{A}_{HF}) + (G^{R}_{HF}-G^{A}_{HF}) \Sigma^{A} \tilde{G}^{A} \\
+ G^{R}_{HF} (\Sigma^{R}-\Sigma^{A}) \tilde{G}^{A} + G^{R}_{HF} \Sigma^{R} (\tilde{G}^{R}-\tilde{G}^{A}).
\end{eqnarray}
By cancellation of some terms we obtain
\begin{eqnarray}
G^{>}-G^{<}=G^{R}_{HF}-G^{A}_{HF}-G^{A}_{HF}\Sigma^{A} \tilde{G}^{A} + G^{R}_{HF} \Sigma^{R} \tilde{G}^{R}=G^{R}-G^{A},
\end{eqnarray}
where we used Eq. \eqref{gkbagra}. Thus, we have consistently extended the GKBA idea to the KBE double time domain.

\section{Equilibrium correlated state from eGKBA}
The eGKBA approximation presented in the previous section for the KBE double-time domain can be also used 
in the equilibrium, see Fig.\ref{chart} iii). Since the eGKBA is defined for the full KBE, we follow the same logic as in the 
KBE case. Then the crucial equation is
\begin{equation}
G^{>}(t,t')-G^{<}(t,t')=G^{R}(t,t')-G^{A}(t,t'),
\end{equation}
which must be fulfilled at all times, (this has been verified in the previous section). 
Assuming that in the long time limit the functions depend only on the time difference,
\begin{equation}
G^{>}(t-t')-G^{<}(t-t')=G^{R}(t-t')-G^{A}(t-t'),
\end{equation}
we can use the Fourier transform to express the equation in the omega space
\begin{equation}
{G}^{>}(\omega)-{G}^{<}(\omega)={G}^{R}(\omega)-{G}^{A}(\omega).
\end{equation}

Here we can imagine that we have arrived to the steady state by the adiabatic switching procedure.
If we further assume that except for the adiabatic switching there is no external force during the evolution 
then after the evolution the adiabatically prepared state is the equilibrium \cite{Keldysh}. Then the equilibrium
state can be represented by the Green's function (on the Matsubara imaginary time segment) which should fulfill
the Kubo--Martin--Schwinger conditions \cite{StefLeeu} and which can be analytically continued to real times.
The analytically continued Green's function satisfy the same boundary conditions \cite{StefLeeu}. Then 
we can finally express the lesser Green's function as 
 \begin{equation}
{G}^{<}_{\rm eq.}(\omega)=-2i{\rm Im}{G}^{R}_{\rm eq.}(\omega)f(\omega).
\end{equation}
This relation is known as fluctuation-dissipation theorem. The density is given by 
 \begin{equation}\label{eqrho}
{\rho}_{\rm eq.}= \int \frac{d\omega}{2\pi} (-i){G}^{<}_{\rm eq.}(\omega)=-\int \frac{d\omega}{\pi}{\rm Im}{G}^{R}_{\rm eq.}(\omega)f(\omega),
\end{equation}
and it should correspond to the density reached by the preparation of the correlated equilibrium state by 
the adiabatic switching $\rho_{HF}\rightarrow\rho_{\rm eq.}$. \\

At this point we have to provide the equation for retarded Green's function, which is an analog of the usual 
Dyson equation.  In the eGKBA approximation the retarded Green's function is computed via Eq. \eqref{gkbagra}, which
becomes in the equilibrium (omitting the equilibrium index)
\begin{equation}\label{equilibrium stategkba}
{G}^{R}(\omega)={G}^{R}_{HF}(\omega) + {G}^{R}_{HF}(\omega) \Sigma^{R}[\tilde{{G}}](\omega) \tilde{{G}}^{R}(\omega).
\end{equation}
This rather interesting equation does not have the standard Dyson structure, as $G^{R/A}$ does not appear on the right hand site. 
The resulting Green's function is not guaranteed to have a positive spectral function, even if $\Sigma$ generates the positive spectral
function in the standard full KBE. This is a pure artefact of the GKBA which focus exclusively on the density matrix, i.e. 
the integral of the spectral function. The mentioned analytic behavior of the retarded Green's function can lead to practical problems
during the iteration of the equilibrium state equations due to the numerically delicate evaluation of $\rho$ from the Eq. \eqref{eqrho}.
In the next section we discuss an approximation of Eq. \eqref{equilibrium stategkba} to avoid the mentioned
practical problems. 

\section{Approximate equilibrium correlated state from eGKBA}

Here we are mainly interested in closed systems where the auxiliary Green's function is usually chosen to be 
the Hartree--Fock Green's function $\tilde{{G}}^{R}(\omega)={G}^{R}_{HF}(\omega)$. Then Eq. \eqref{equilibrium stategkba}
becomes
\begin{equation}\label{gkbareteq}
{G}^{R}(\omega)={G}^{R}_{HF}(\omega) + {G}^{R}_{HF}(\omega) \Sigma^{R}[{G}_{HF}](\omega) {G}^{R}_{HF}(\omega).
\end{equation}
This equation can still be problematic to solve, since the spectral features mentioned above persist. 
However, we can make an observation about an approximate retarded Green's function given by
\begin{equation}\label{approximativesol}
{G}^{R}_{\rm appr.} = {G}^{R}_{HF} + {G}^{R}_{HF} \Sigma^{R} {G}^{R}_{\rm appr.} = {G}^{R}_{HF} + {G}^{R}_{HF} \Sigma^{R} {G}^{R}_{HF}
+ {G}^{R}_{HF} \Sigma^{R} {G}^{R}_{HF} \Sigma^{R} {G}^{R}_{HF}  \dots
\end{equation}
If the second and higher orders of the expansion are negligible, the approximate Green's function will be the same as in Eq. (\ref{gkbareteq}). The 
advantage of this equation is that it has the Dyson structure and gives a numerically less delicate evaluation of $\rho$. 
The error between the approximate Green's function and the Green's function 
\begin{equation}
G^{R}_{\rm appr.}-G^{R}_{}={G}^{R}_{HF} \Sigma^{R} {G}^{R}_{HF} \Sigma^{R} {G}^{R}_{HF}+\dots
\label{hybnew13}
\end{equation}
is of the second order in the interaction expansion and it grows with the interaction strength.\\

\section{Approximate equilibrium correlated state from eGKBA - example}

\begin{figure}[t!]
\centering
\resizebox{0.85\columnwidth}{!}{%
  \includegraphics{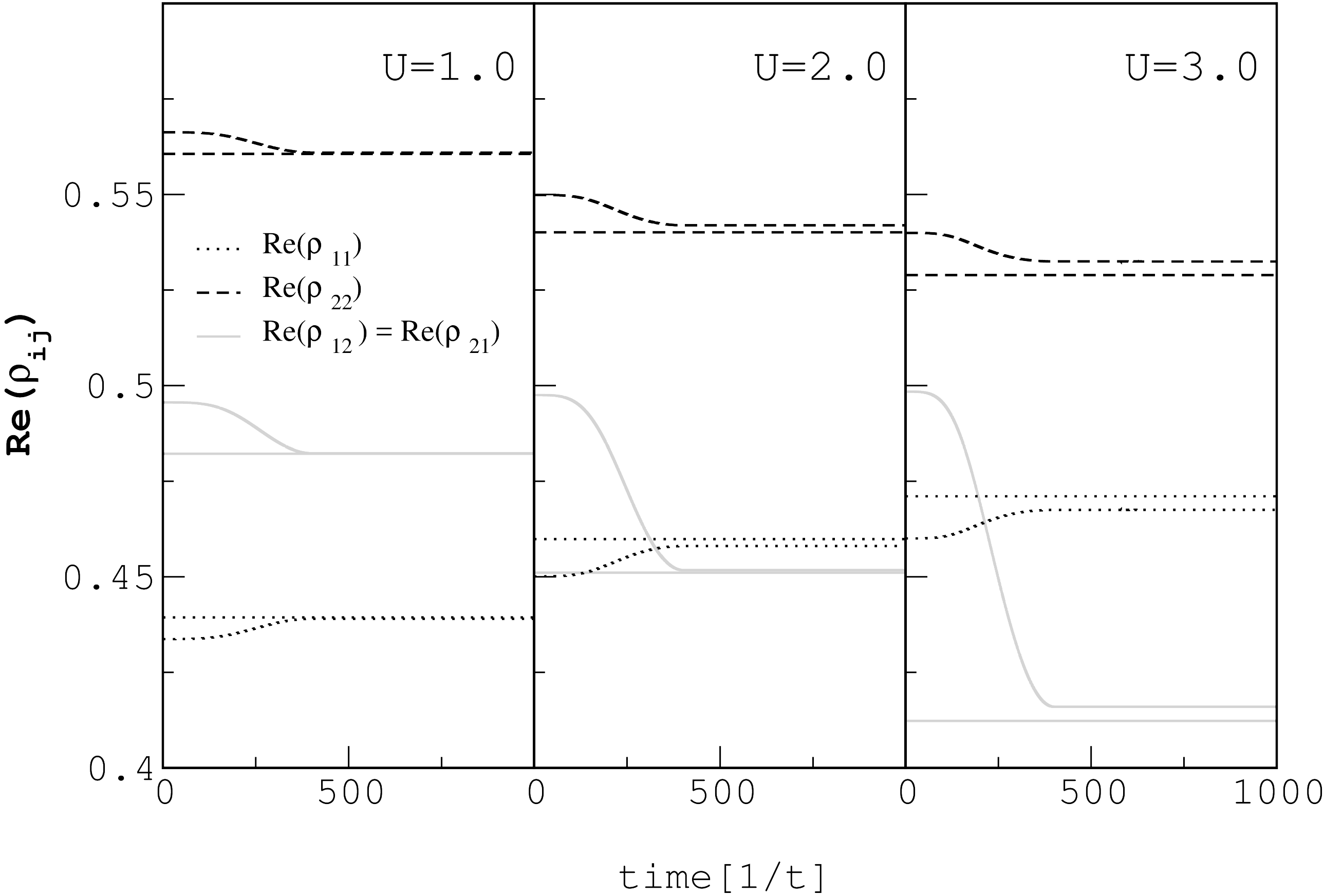} }
\caption{The density matrix elements for the Hubbard dimer $v_{1}-v_{2}=0.4t$ where $t=1$ is the 
hopping and $v_{i}$ are the on-site energies, for the different strengths of the on-site interactions $U_{1}=U_{2}=U$. 
Time is measured in inverse of the hopping $t$. A comparison of the GKBA adiabatic switching with the approximate 
correlated equilibrium state (straight lines).}
\label{fig:1}
\end{figure}

The solution of the approximate correlated equilibrium (ground) state is illustrated using a Hubbard dimer, i.e. a  
two-site, tight-binding cluster, with one orbital per site: 
\begin{equation}
{H}_{\rm dimer}=\sum_{\sigma}\bigl[-t(c_{1\sigma}^{\dagger}c_{2\sigma}^{}+c.c.)+{v}_{1}n_{1\sigma}+{v}_{2}n_{2\sigma}\bigr]+{U}_{1}n_{1\uparrow}n_{1\downarrow}+{U}_{2}n_{2\uparrow}n_{2\downarrow}.
\label{ham}
\end{equation}
This system, is exactly solvable. Here we consider the case of two electrons with opposite spin projections, which
mutually interact when at the same site. \newline

We solve ${G}^{R}_{\rm appr.}$  given by the Eq. \eqref{eqrho} and Eq. \eqref{approximativesol} with the Second-Born correlation self-energy
which, written in the orbital indexes $i$ and $j$, reads
\begin{equation}
\begin{split}
&\Sigma^{R}_{ij}(\omega)=U_{i}U_{j}\int\int\frac{d\omega' d\omega''}{(2\pi)^2}\Bigl(\tilde{G}^{R}_{ij}(\omega')\tilde{G}^{<}_{ji}(\omega'')\tilde{G}^{<}_{ij}(\omega-\omega'+\omega'')+\\
&+\tilde{G}^{R}_{ij}(\omega')\tilde{G}^{<}_{ji}(\omega'')\tilde{G}^{R}_{ij}(\omega-\omega'+\omega'')-\tilde{G}^{R}_{ij}(\omega')\tilde{G}^{<}_{ji}(\omega''){(\tilde{G}^{R}_{ji})}^{*}(\omega-\omega'+\omega'')+\\
&+\tilde{G}^{<}_{ij}(\omega'){(\tilde{G}^{R}_{ij})}^{*}(\omega'')\tilde{G}^{<}_{ij}(\omega-\omega'+\omega'')+\tilde{G}^{<}_{ij}(\omega')\tilde{G}^{<}_{ji}(\omega'')\tilde{G}^{R}_{ij}(\omega-\omega'+\omega'')\Bigr),
\end{split}
\end{equation}
where, as in the GKBA,  the lesser and greater function are
\begin{eqnarray}
&\tilde{G}^{<}_{ij}(\omega)=\sum_{l}\bigl[-\tilde{G}^{R}_{il}(\omega)\rho_{lj}+\rho_{il}\tilde{G}^{A}_{lj}(\omega)\bigr],\\
&\tilde{G}^{>}_{ij}(\omega)=\sum_{l}\bigl[\tilde{G}^{R}_{il}(\omega)(1-\rho)_{lj}-(1-\rho)_{il}\tilde{G}^{A}_{lj}(\omega)].
\end{eqnarray}
and $\tilde{{G}}^{R/A}={G}^{R/A}_{HF}$  is used as the auxiliary GKBA Green's function.\\

In Fig. \ref{fig:1} we show a comparison of the direct solution with the GKBA adiabatic switching $\rho_{HF}\rightarrow\rho_{\rm eq.}$. 
From the results we can clearly see that for strong enough interactions there is a deviation, since
we compute the equilibrium state only approximately. Here, the higher order terms given by \eqref{hybnew13} are not negligible.
However, for lower interaction strengths where we can neglect these terms, the approximation is remarkably good.

\section{Conclusions}

We have shown how to directly obtain the initial equilibrium state for GKBA approximation, thus avoiding 
the usual adiabatic switching protocol. For this purpose we have extended the GKBA approximation from the transport equation
for $\rho$ to the full Kadanoff--Baym equation. The extended GKBA was then used to define the equilibrium-state equation.
The approximate solution of the extended GKBA was compared to the adiabatic switching procedure 
for a case of Hubbard dimer with good results. Our work provides an alternative way how to find the initial equilibrium
GKBA states and, at the same time, provides a new and broader perspective on the GKBA approximation. 
It would be interesting to test eGKBA on other (and with more degrees of freedom) systems, and for approximations other than Second-Born, and 
this is left as future work.\\

{\footnotesize
Author contribution: C.V. conceived the idea and supervised the project; M.H. developed the theoretical 
formulation and performed the numerical simulations. Both authors contributed to the manuscript.

\end{document}